\documentclass[english,12pt,aps,prd,a4paper,preprintnumbers,floatfix,nofootinbib,showpacs,superscriptaddress, notitlepage]{revtex4-1} 
\pdfoutput=1
\usepackage{color}
\usepackage{graphicx}
\usepackage{caption}
\usepackage{subcaption}
\usepackage{amsmath}
\usepackage{amssymb}
\usepackage[colorlinks=true,citecolor=darkred,urlcolor=darkred, pdfborder={0 0 0}]{hyperref}
\usepackage[normalem]{ulem}

\definecolor{darkred}{rgb}{0.6,0,0}

\definecolor{linkcolor}{rgb}{0,0,0.5}
\usepackage{dcolumn}% Align table columns on decimal point
\usepackage{bm}% bold math
\usepackage{graphicx}
\usepackage{multirow}
\usepackage{rotating}
\usepackage{subcaption}

\usepackage{epstopdf}
\usepackage[utf8]{inputenc}

\def\gsim{\raise0.3ex\hbox{$\;>$\kern-0.75em\raise-1.1ex\hbox{$\sim\;$}}}
\def\lsim{\raise0.3ex\hbox{$\;<$\kern-0.75em\raise-1.1ex\hbox{$\sim\;$}}}

\def\beqn#1{\begin{equation}\label{#1}}
\def\eeqn{\end{equation}}

\def\beqa#1{\begin{eqnarray}\label{#1}}
\def\eeqa{\end{eqnarray}}

\def\21{$\mathrm{SU(2)_L \otimes U(1)_Y}$ }
\def\31{$\mathrm{SU(3)_c \otimes U(1)_Q}$ }

\def\3211{$\mathrm{SU(3) \otimes SU(2)_L \otimes U(1)_R \otimes U(1)_{B-L}}$ }
\def\321{$\mathrm{SU(3) \otimes SU(2) \otimes U(1)}$ }
\def\422{$\mathrm{SU(4) \otimes SU(2) \otimes SU(2)_R}$ }
\newcommand {\ignore}[1]{}

\newcommand{\sm}{{Standard Model }}

\newcommand{\AddrAHEP}{%
  AHEP Group, Institut de F\'{i}sica Corpuscular --
  CSIC-Universitat de Val\`{e}ncia, Parc Cient\'ific de Paterna.\\
 C/ Catedr\'atico Jos\'e Beltr\'an, 2 E-46980 Paterna (Valencia) - SPAIN}
 
\newcommand{\AddrUNAM}{ {\it Instituto de F\'{\i}sica, Universidad Nacional Aut\'onoma de M\'exico, A.P. 20-364, Ciudad de M\'exico 01000, M\'exico.}}

\begin{document}

\bibliographystyle{unsrt}   % needed for refs and hyperlinks %% utphys.bst style file is also needed with this %%%%%%%%%%%%

\title{Scotogenic Dark Symmetry as a residual subgroup of Standard Model Symmetries}

\author{Salvador Centelles Chuli\'{a}}\email{salcen@ific.uv.es}
\affiliation{\AddrAHEP}
\author{Ricardo Cepedello}\email{ricepe@ific.uv.es}
\affiliation{\AddrAHEP}
\author{Eduardo Peinado}\email{epeinado@fisica.unam.mx}
\affiliation{\AddrUNAM}
\author{Rahul Srivastava}\email{rahulsri@ific.uv.es}
\affiliation{\AddrAHEP}

\begin{abstract}
  \vspace{1cm} 
We show that the scotogenic dark symmetry can be obtained as a residual subgroup of the global $U(1)_{B-L}$ symmetry already present in Standard Model. We propose a general framework where the  $U(1)_{B-L}$ symmetry is spontaneously broken to an even $\mathcal{Z}_{2n}$ subgroup, setting the general conditions for neutrinos to be Majorana and the dark matter stability in terms of the residual $\mathcal{Z}_{2n}$. Under this general framework, as examples, we build a class of simple models where, in the scotogenic spirit, the dark matter candidate is the lightest particle running inside the neutrino mass loop.
The global $U(1)_{B-L}$ symmetry in our framework being anomaly free can also be gauged in a straightforward manner leading to a richer phenomenology.
   
\end{abstract}

\maketitle

%%%%%%%%%%%%%%%%%%%%%%%%%%%%%%%%%%%%%%%%%%%%%%%%%%%%%%%%%%%%%%%%%%%%%%%%
\section{Introduction}
%%%%%%%%%%%%%%%%%%%%%%%%%%%%%%%%%%%%%%%%%%%%%%%%%%%%%%%%%%%%%%%%%%%%%%%%

The Standard Model of particle physics is a highly successful theory with an enormous predictive power.
So far it has passed each and every experimental scrutiny with flying colors and many of its predictions have been experimentally verified.
The discovery of a scalar boson in 2012 at the LHC \cite{Chatrchyan:2012xdj, Aad:2012tfa}, if confirmed to be the \sm Higgs, will be the icing on the cake.
With hundreds of precision observables and dozens of predicted particles, it is without doubt one of the most precise theories in the history of human science.
Despite its success, the \sm also has some serious drawbacks that need to be addressed in order to obtain a more complete fundamental theory. Two of the main issues it faces, although not the only ones, are neutrino masses and dark matter.

Although dark matter direct detection experiments so far have shown only negative results \cite{Akerib:2016vxi, Aprile:2018dbl}, the cosmological evidences for its existence are abundant.
Observations ranging from the galaxy rotation curves to galaxy clusters or gravitational lensing, all point to the existence of dark matter, a hitherto unknown type of matter which interacts gravitationally but has little to no electromagnetic interaction \cite{Aghanim:2018eyx}. 
From a particle-physics point of view, a completely stable or sufficiently long lived, electrically neutral but weakly interacting massive particle (WIMP) is one of the most popular candidates for dark matter. 
The \sm unfortunately has no such candidate for dark matter.
This creates the need to extend the matter content and possibly the symmetry inventory to explain the cosmological observations pointing towards the existence of dark matter.
In this letter, we show that the global $U(1)_{B-L}$ already present in the \sm is enough to ensure the stability of dark matter. 
Furthermore, such a dark matter candidate can be intimately related with the other major experimental shortcoming of the Standard Model, namely the lack of a neutrino mass generation
mechanism.

The neutrinos are predicted to be massless in the Standard Model.
However, thanks to data from various oscillation experiments, we can confidently say that neutrinos are massive particles \cite{McDonald:2016ixn,Kajita:2016cak,An:2012eh,Abe:2017uxa,deSalas:2018bym, deSalas:2017kay}. 
Consequently, the \sm has to be extended in one way or another to accommodate massive neutrinos \cite{Ma:1998dn, CentellesChulia:2018gwr}.
In the past several decades, various extensions of the \sm have been proposed to understand massive neutrinos.
Most of the first works on neutrino mass models assumed that neutrinos are Majorana and proposed several seesaw \cite{Minkowski:1977sc,Yanagida:1979as,Mohapatra:1979ia,Schechter:1981cv, Schechter:1980gr,Foot:1988aq} and loop mass mechanisms \cite{Zee:1980ai, Babu:1988ki, Ma:2006km} to explain their small yet non-zero masses.
Majorana neutrino mass models still remain the popular choice for the \sm extensions that try to explain massive neutrinos \cite{Bonnet:2012kz, Ma:2013mga, Sierra:2014rxa,Ma:2015xla, CarcamoHernandez:2016pdu, Cepedello:2017eqf, Bernal:2017xat, Cepedello:2018rfh, Anamiati:2018cuq,CarcamoHernandez:2019cbd}.  

Typically, Majorana mass models break the global Lepton number $U(1)_L$ symmetry (or equivalently, the anomaly free $U(1)_{B-L}$ symmetry) of the \sm to a residual $\mathcal{Z}_2$ subgroup.
However, breaking $U(1)_{B-L}$ to higher $\mathcal{Z}_m$ subgroups is also feasible, where $m\in \mathbb{Z}^+$ and $m \geq 2$, $\mathbb{Z}^+$ being set of all positive integers. 
In fact, in absence of any other conserved symmetries beyond the Standard Model, the Dirac or Majorana nature of neutrinos depends on the $U(1)_{B-L}$
breaking pattern as argued in \cite{Hirsch:2017col,Bonilla:2018ynb}.
If the  $U(1)_{B-L}$ remains conserved, then neutrinos have to be Dirac, as the Majorana mass term is forbidden by it. 
In the case where $U(1)_{B-L}$ breaks to a residual $\mathcal{Z}_m$ subgroup with the \sm lepton doublets $L_i = (\nu_{L_i}, l_{L_i})^T$; $i = 1,2,3$ transforming non-trivially under it, then \cite{Hirsch:2017col,Bonilla:2018ynb}
\begin{eqnarray} \label{oddzn}
U(1)_{B-L}   & \, \to  \, &   \mathcal{Z}_m \equiv \mathcal{Z}_{2n+1} \, \text{with} \,  n \in \mathbb{Z}^+    \nonumber \\
& \, \Rightarrow \, & \text{Neutrinos are Dirac particles}      \nonumber \\
U(1)_{B-L}  & \, \to  \,  & \mathcal{Z}_m \equiv \mathcal{Z}_{2n} \, \text{with} \,  n \in \mathbb{Z}^+  \\
& \, \Rightarrow \, & \text{Neutrinos can be Dirac or Majorana } \nonumber
%  \nonumber \\
\end{eqnarray}
For the case when $U(1)_{B-L}$ is broken down to an even $\mathcal{Z}_{2n}$ residual subgroup, one can make a further classification depending on how $L_i$ transform under $\mathcal{Z}_{2n}$,
\begin{eqnarray} \label{evenzndir}
 L_i \left\{ \begin{array}{ll}
          \nsim  \omega^{n} \ \ \text{under $\mathcal{Z}_{2n}$} &  \Rightarrow\text{Dirac Neutrinos}\\
          \sim  \omega^{n}\ \ \text{under $\mathcal{Z}_{2n}$} &  \Rightarrow\text{Majorana Neutrinos}\end{array} \right. 
\end{eqnarray}
where $ \omega = e^{\frac{2\pi i}{2n}}$ is the $\rm{2n}^{\rm{th}}$ root of unity with  $\omega^{2n}=1$. 
Thus, one can obtain Majorana neutrinos also in cases when $U(1)_{B-L}$ is broken to any even $\mathcal{Z}_{2n}$ residual subgroup beyond $\mathcal{Z}_2$.
Despite a very large amount of literature on mass mechanisms for Majorana neutrinos, to best of our knowledge,
the option of $U(1)_{B-L} \to \mathcal{Z}_{2n}$; $n \geq 2$ leading to Majorana neutrinos in congruence with \eqref{evenzndir} has not been explored.

In this letter, we show that for Majorana neutrinos with $U(1)_{B-L} \to \mathcal{Z}_{2n}$, the residual $\mathcal{Z}_{2n}$ subgroup can also be used to obtain a stable candidate for dark matter without adding any new symmetry to the Standard Model.
We further show that such a dark matter candidate can also be intimately connected with the neutrino mass generation mechanism, being the lightest of the loop mediators leading to Majorana masses for the neutrinos \textit{a la scotogenic} \cite{Ma:2006km}. 
Thus, the residual $\mathcal{Z}_{2n}$ subgroup plays the role of scotogenic dark symmetry, which in the original scotogenic model had to be imposed as an extra ad hoc condition.\footnote{For an interpretation of the scotogenic dark symmetry as matter parity, see \cite{Ma:2015xla}.}

We will start our discussion in Section \ref{method} highlighting the general conditions required to have a loop mass generation mechanism for neutrinos with the residual $\mathcal{Z}_{2n}$ symmetry playing the role of the scotogenic dark symmetry.
After illustrating the general applicability of the framework, we will discuss a class of simple one-loop models that can be constructed using our general criterion.
In Section \ref{model} we will take one of these models as an explicit example and discuss it in further details.
We will finally conclude our discussion and summarize the main results in Section \ref{summary}.

%%%%%%%%%%%%%%%%%%%%%%%%%%%%%%%%%%%%%%%%%%%%%%%%%%%%%%%%%%%%%%%%%%%%%%%%
\section{The General Formalism} \label{method}
%%%%%%%%%%%%%%%%%%%%%%%%%%%%%%%%%%%%%%%%%%%%%%%%%%%%%%%%%%%%%%%%%%%%%%%%

As pointed out before in \eqref{oddzn} and \eqref{evenzndir}, in order to have Majorana neutrinos one has to break $U(1)_{B-L}$ symmetry into an even subgroup $\mathcal{Z}_{2n}$.
In addition, the lepton doublets $L_i$ should also belong to the subgroup $\mathcal{Z}_2 \subset \mathcal{Z}_{2n}$, i.e. $L_i$ either transform trivially or as $\omega^n$ with $\omega^{2n}=1$. 
A connection between these symmetries and the stability of dark matter can be found, as first stated in \cite{Bonilla:2018ynb} for Dirac neutrinos.
In this letter, we follow an analogous approach linking the generation of naturally small Majorana neutrino masses with the stability of dark matter providing the appropriate symmetry breaking pattern $U(1)_{B-L}\rightarrow\mathcal{Z}_{2n}$. This further implies that neutrino masses arise at loop level as the tree-level Majorana and Dirac masses
are forbidden by the symmetry.

\begin{figure}[th!]
\hspace*{-0.5cm}
    \centering
    \begin{subfigure}[t]{0.5\textwidth}
        \centering
        \includegraphics[width=1\textwidth]{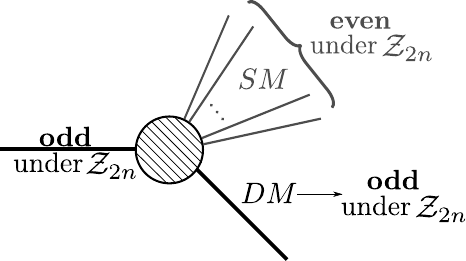}
        \caption{The lightest of the odd fields under $\mathcal{Z}_{2n}$ will be the dark matter candidate.}
    \end{subfigure}%
    ~ 
    \begin{subfigure}[t]{0.55\textwidth}
        \centering
        \includegraphics[width=0.55\textwidth]{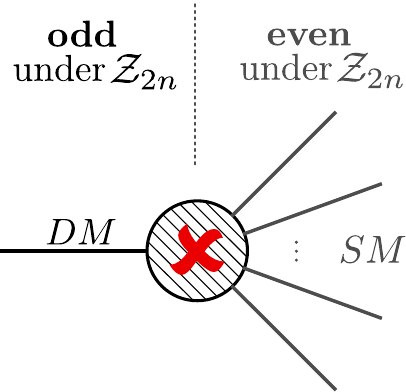}
        \caption{The decay of the dark matter to the Standard Model is forbidden by the residual symmetry.}
    \end{subfigure}
    \caption{There are two distinct sectors transforming as \textit{odd} or \textit{even} under the residual $\mathcal{Z}_{2n}$ symmetry. In our setup, all the internal fields are odd, while the SM is even. Due to the $\mathcal{Z}_{2n}$ symmetry, an odd particle can only decay to the SM plus another odd particle. Thus, the lightest of the odd particles is stable and a good dark matter candidate.}
    \label{U1mod}
\end{figure}

In order to do this, new fields with exotic $B-L$ charges are required.\footnote{Note that every fermion has to be massive, this means that they should have a vector-like partner or if chiral, one should link the breaking of $U(1)_{B-L}$ to their mass generation \cite{Bonilla:2018ynb}.}
Since in the \sm lepton doublets $L_i$ have $B-L$ charge $-1$, in order to avoid all possible tree-level Dirac mass terms, no new fermion can carry $\pm 1$ charges under $U(1)_{B-L}$ symmetry.
Furthermore, the lowest order Majorana mass term, i.e. the Weinberg operator $\bar{L}^cLHH$, is not invariant under $U(1)_{B-L}$, so it is automatically absent.
To generate neutrino masses we should go to higher dimensional operators,
\begin{equation} \label{eq:operator}
    \bar{L}^cLHH \chi_1 ...\chi_k,
\end{equation}
where the $\chi_i$; $i = 1, ... k$  are scalars fields transforming non-trivially under $U(1)_{B-L}$. 
The operator in \eqref{eq:operator} should be invariant under the Standard Model symmetries including $U(1)_{B-L}$.
This means that the $B-L$ charges of the fields $\chi_i$ must sum up to $2$.
Although in principle some of them can also have non-trivial transformations under $SU(2)_L \otimes U(1)_Y$, for sake of simplicity we will take all $\chi_i$ to be \sm gauge singlets.
Since the $\chi_i$ are charged under that $U(1)_{B-L}$, once they acquire a vacuum expectation value (vev), the $U(1)_{B-L}$ symmetry will break down to a residual $\mathcal{Z}_{2n}$ subgroup, with $n$ depending on the charges of the particles in the model. 

As has been pointed out previously in \cite{Chulia:2016ngi, CentellesChulia:2018gwr, Bonilla:2018ynb}, the stability of dark matter can be achieved automatically if $U(1)_{B-L}$ is broken to an even $\mathcal{Z}_{2n}$ provided that all the Standard Model particles transform as \textit{even} under $\mathcal{Z}_{2n}$, while the dark matter candidate is \textit{odd}. Here by \textit{even} (\textit{odd}) we mean fields which transform as even (odd) powers of $\omega$ under $\mathcal{Z}_{2n}$ with $\omega^{2n}=1$.
Thus, the necessity of the $U(1)_{B-L}$ breaking to an even $\mathcal{Z}_{2n}$ does not only come from the Majorana nature of neutrinos but also from the requirement to have a stable dark matter in this setup.
An even residual $\mathcal{Z}_{2n}$ symmetry ensures that if the Standard Model belongs to the $\mathcal{Z}_n$ subgroup of it, then a \textit{dark sector} with  all the fields transforming as \textit{odd} under $\mathcal{Z}_{2n}$, is separated from it.
The interplay between both sectors and the stability of dark matter can be seen graphically in Figure \ref{U1mod}. Note that any particle odd under the residual $\mathcal{Z}_{2n}$ symmetry can only decay into the Standard Model particles plus another odd particle. This implies that the lightest of the odd particles will be automatically stable, see \cite{Bonilla:2018ynb} for more details.

\subsection*{One-loop realizations of the operator $\bar{L}^c LHH \chi$}

\begin{figure}[th!]
    \centering
      \hspace*{1cm}  \includegraphics[width=0.89\textwidth]{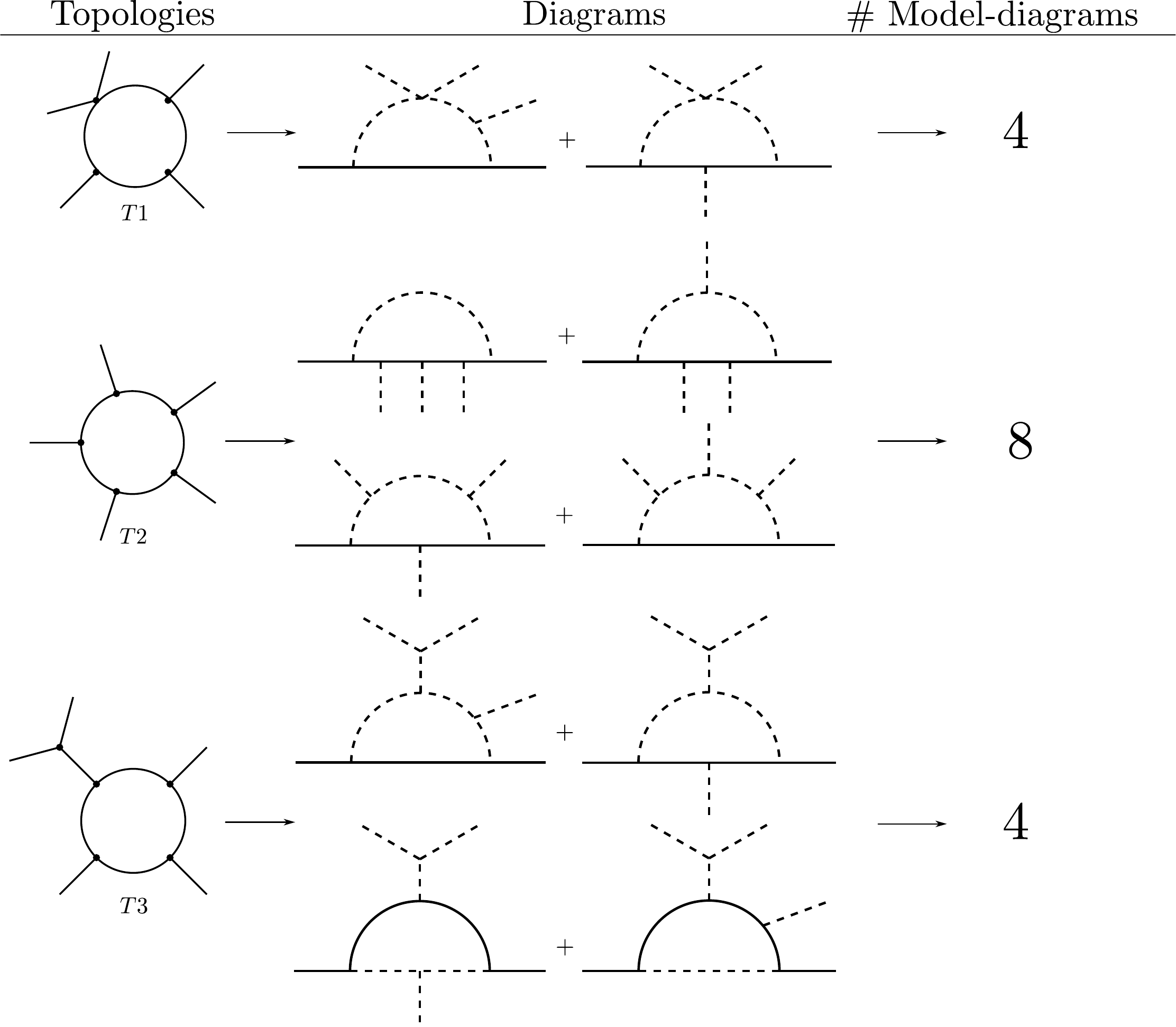}
    \caption{Renormalizable genuine topologies that generate the operator $\bar{L}^c LHH \chi$. For each topology, all the diagrams are given along with the number of model-diagrams. Each model-diagram can be generated by arranging in all possible ways the $\chi$ and the two $H$ in the external scalar legs.}
    \label{fig:topos}
\end{figure}

Following this framework, in the most simple scenario, one can realize the operator \eqref{eq:operator} at the one-loop level with only one field $\chi$ with $B-L$ charge $2$, i.e. the dimension 6 operator $\bar{L}^c LHH \chi$. The possible one-loop realizations of the operator can be classified, following the philosophy of \cite{Bonnet:2012kz, Sierra:2014rxa,Cepedello:2018rfh, Anamiati:2018cuq, Cepedello:2017eqf}, into three renormalizable genuine topologies which lead to 10 different diagrams as can be seen in Figure~\ref{fig:topos}. We associate topologies with graphs or Feynman diagrams where no Lorentz nature is considered. We refer to diagrams if fermion and scalar lines are specified. The concept of genuinity is then attributed to those models for which the main contribution to neutrino masses comes from the one-loop level realization of the operator $\bar{L}^c LHH \chi$. We call topologies or diagrams that generate at least one of these models genuine by inference. For example, diagrams which unavoidably contain the vertices ($L H$ $+$ fermion) or ($\bar{L}^c L$ $+$ scalar) are not genuine as they would generate a dominant type-I/III or type-II seesaw contribution, respectively.
 
The ten different diagrams depicted in Figure~\ref{fig:topos} generate 18 model-diagrams. Each model-diagram is generated from a given diagram by the different arrangements of the two Higgs doublets and the Higgs singlet $\chi$ of $\bar{L}^c LHH \chi$ in the external scalar lines. For instance, take topology $T1$, each of its diagrams generate two model-diagrams inserting $\chi$: (1) in the quartic scalar coupling or (2) in the trilinear coupling with scalars or fermions. In the case of $T3$ the arrangement of $\chi$ and both Higgses is unique, as a trilinear vertex with two $H$ is not allowed because it would generate a dominant type-II seesaw contribution. Note that for each model-diagram there is an infinite series of possible models as there is always a free set of charges running in the loop. 

The intention of this letter is not to make an exhaustive classification, but to show in a systematic way the wide range of possibilities, yet unexplored, of the most simple realizations of the framework given in this section. 
We will now choose one of the simplest diagrams to build a particular, consistent and complete model as an example of how this general method works.

%%%%%%%%%%%%%%%%%%%%%%%%%%%%%%%%%%%%%%%%%%%%%%%%%%%%%%%%%%%%%%%%%%%%%%%%
\section{A simple explicit model} \label{model}
%%%%%%%%%%%%%%%%%%%%%%%%%%%%%%%%%%%%%%%%%%%%%%%%%%%%%%%%%%%%%%%%%%%%%%%%

In this section, we construct an explicit Ultra-Violet (UV) complete model realization of the dimension 6 operator $\bar{L}^c LHH \chi$, in order to further describe the formalism developed in Section \ref{method}. 
We add a new vector-like fermion pair $F_L$ and $F_R$ with charge $1/2$ under $U(1)_{B-L}$ but singlet under the \sm gauge symmetries.
Since the field breaking the $U(1)_{B-L} $ symmetry, $\chi$, transforms as $2$, the fractional charges of the new fields will imply the breaking pattern is $U(1)_{B-L} \rightarrow \mathcal{Z}_4$.
Note that, given the fractional charges of $F_L$ and $F_R$, there will be no tree level Dirac mass term for neutrinos. 
Thus, additional scalars $\eta_i$; $i =1,2,3$ are also needed to generate a one-loop contribution to neutrino masses.
The relevant matter fields and their transformation under $SU(2)_L \otimes U(1)_Y\otimes U(1)_{B-L}$ are given in Table \ref{tab1}, as well as the charges under the residual  $\mathcal{Z}_4$ subgroup that survives after spontaneous symmetry breaking.

\begin{table}
\begin{center}
\begin{tabular}{| c || c | c | c | c |}
  \hline 
&   \hspace{0.1cm}  Fields     \hspace{0.1cm}       &    $SU(2)_L \otimes U(1)_Y$            &    \hspace{0.2cm}   $U(1)_{B-L}$        \hspace{0.2cm}               & 
   \hspace{0.4cm} $\mathcal{Z}_{4}$ \hspace{0.4cm}                             \\
\hline \hline
\multirow{4}{*}{ \begin{turn}{90} Fermions \end{turn} } &
 $L_i$        	  &    ($\mathbf{2}, {-1/2}$)       &   {\color{red}$-1$}    	  &	 {\color{blue}$\omega^2$}                     \\	
&   $e_{R_i}$       &   ($\mathbf{1}, {-1}$)      & {\color{red} $-1$}   &  	 {\color{blue}$\omega^2$}\\
&   $F_{R}$       &   ($\mathbf{1}, {0}$)      & {\color{red} $1/2$}   &  	 {\color{blue}$\omega$}\\
&   $F_{L}$    	  &   ($\mathbf{1}, {0}$)      & {\color{red}$1/2$}   &    {\color{blue}$\omega$}     \\
\hline \hline
\multirow{5}{*}{ \begin{turn}{90} Scalars \end{turn} } &
 $H$  		 &  ($\mathbf{2}, {1/2}$)      &  {\color{red}$0$}    & {\color{blue}$1$}    \\
& $\chi$          	 &  ($\mathbf{1}, {0}$)        &  {\color{red}$2$ }  &  {\color{blue} $1$}     \\		
& $\eta_1$          	 &  ($\mathbf{2}, {-1/2}$)      &  {\color{red}${-3/2}$}    &  {\color{blue}$\omega$}       \\
& $\eta_2$             &  ($\mathbf{2}, {-1/2}$)        &  {\color{red}${-1/2}$}      &	{\color{blue}$\omega^3$} \\
& $\eta_3$             &  ($\mathbf{2}, {-1/2}$)        &  {\color{red}${3/2}$}      &	{\color{blue}$\omega^3$} \\	
    \hline
  \end{tabular}
\end{center}
\caption{Particle content of the model with $i\in\{e, \mu, \tau\}$. All the fields listed in the table are $SU(3)_C$ singlets. The field $\chi$ acquires a vev breaking consequently the $U(1)_{B-L}$ symmetry into its $\mathcal{Z}_4$ subgroup given the half-integer charges running in the loop (see text for details).}
 \label{tab1}
\end{table}%

It is clear that the $U(1)_{B-L}$ symmetry given in Table \ref{tab1} is anomalous. The canonical solution to make $U(1)_{B-L}$ anomaly free is to add three right-handed fermions $N_R$ with $(-1,-1,-1)$ charges under $B-L$ symmetry.
However, as noted before in Section \ref{method}, these charges are not allowed as they lead to tree level Dirac coupling between the \sm lepton doublets $L_i$.
Instead to cancel the anomalies, one can simply add three new neutral right-handed fermions $N_R$ with charges $(-4,-4,5)$ under $U(1)_{B-L}$.
This charge assignment also leads to anomaly free $U(1)_{B-L}$ symmetry \cite{Montero:2007cd,Ma:2014qra,Ma:2015mjd,Ma:2015raa}.
Other anomaly free solutions with several additional chiral fermions carrying exotic $B-L$ charges, can also be found as discussed 
in \cite{Ho:2016aye,Ma:2016nnn,Patra:2016ofq, Wang:2016lve, Wang:2017mcy, Nanda:2017bmi, Han:2018zcn, Kang:2018lyy}.
However, the $(-4,-4,5)$ solution seems to be minimal \footnote{Some of the subsequent work on $(-4,-4,5)$ can be found in \cite{Modak:2016ung, Singirala:2017see, DeRomeri:2017oxa,  Nomura:2017jxb, Das:2017deo, Okada:2018tgy, Calle:2018ovc}.}.
These right handed neutrinos can be given Majorana masses through vev of singlet Higgses with charges $8$ and $10$ under $B-L$ \footnote{Note that vev to these Higgses is also consistent with the $U(1)_{B-L} \to \mathcal{Z}_4$ breaking.}.
The  $N_R$ will not play a role in the light neutrino mass generation or in the dark matter phenomenology, but they could be relevant in colliders, particularly if one gauges the $U(1)_{B-L}$ symmetry. 

\begin{figure}[h!t]
    \centering
    \begin{subfigure}[b]{0.38\textwidth}
        \includegraphics[width=\textwidth]{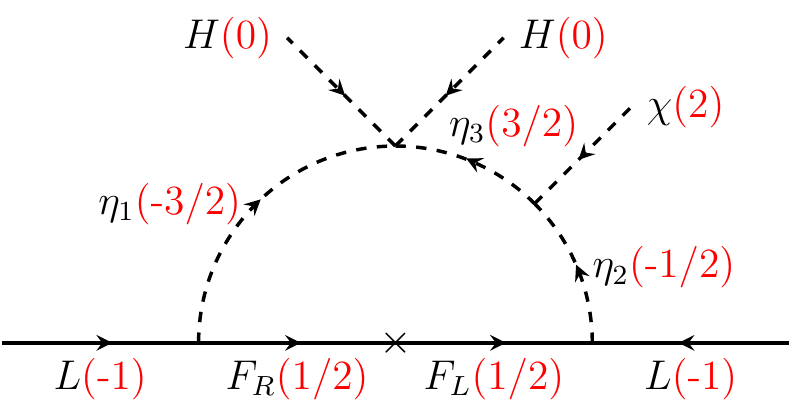}
    \end{subfigure}
    \begin{subfigure}[t]{0.13\textwidth}
        \vspace*{-2.2cm}
        \hspace*{-0.1cm}
        \includegraphics[width=\textwidth]{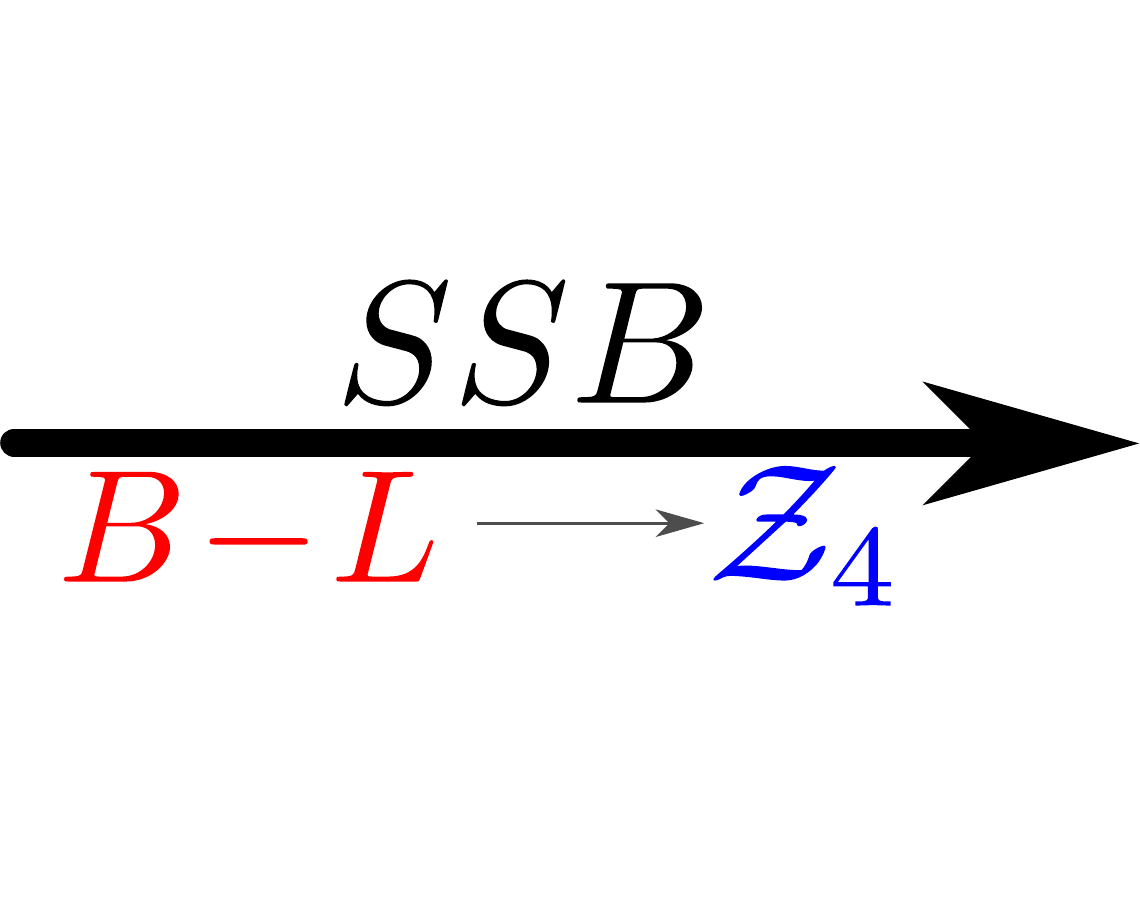}
    \end{subfigure}
    \begin{subfigure}[b]{0.38\textwidth}
        \includegraphics[width=\textwidth]{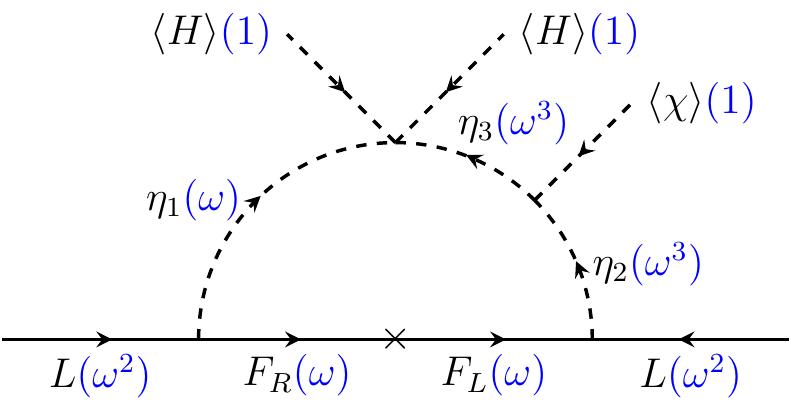}
    \end{subfigure}
    \caption{Leading order neutrino mass diagram with $B-L$ charges (left). After spontaneous symmetry breaking (SSB), as $\chi$ has charge $2$ under $B-L$ and there are half-integer charged fields, $U(1)_{B-L}$ is broken to its subgroup $\mathcal{Z}_4$ (right).}\label{diagloop}
\end{figure}

With this setup, the anomaly free $B-L$ will forbid the tree-level mass term for the neutrinos, but the new field content can accommodate the one-loop neutrino mass diagram of Figure~\ref{diagloop} in the \textit{scotogenic spirit}, thus explaining the smallness of neutrino masses and dark matter stability in a natural way.
We will now write down the complete Lagrangian in several pieces for a better understanding. The Lagrangian of the model consist of the following parts:

\begin{enumerate} 
\item The charged lepton mass Lagrangian is exactly identical to the Standard Model one:
\begin{eqnarray}
\mathcal{L}_{charged} \, = \,Y_l\, \bar{L} \, H^c \, l_R,
\label{lag1}
\end{eqnarray}
where $Y_l$ is a $3 \times 3$ Yukawa matrix and $L$ and $l_R$ are 3-vectors. After spontaneous symmetry breaking $Y_l/\text{v}$ will be the mass matrix for the charged lepton, where v is the vaccum expectation value of the Standard Model Higgs $H$.

\item The mass of the vector-like pair $F_{L}$ and $F_{R}$ will simply be given by:
\begin{equation}
% \mathcal{L}_N = M_{ij} \bar{F}_{L_i} F_{R_j} \, + \, h.c.,
 \mathcal{L}_N = M \bar{F}_{L} F_{R} \, + \, h.c.,
 \label{lag4}
\end{equation}

where $M$ is a general mass matrix whose entries are expected to be much larger than the electroweak scale and therefore $F$ will be a heavy Dirac fermion.  We will use $m_F$  for the eigenvalues of $M$

\item There are also other Yukawa interaction terms that will not contribute to the tree-level fermion masses but appear in the one-loop level:
\begin{eqnarray}
\mathcal{L}_{YukInt} \,= \, Y_1 \, \bar{L} \, F_{R} \, \eta_1 \, + \, Y_2 \, \bar{L}^c \, F_{L} \, \eta_2^\dagger + \,h.c.
\label{lag5}
\end{eqnarray}
\end{enumerate}

Apart from the standard kinetic and gauge terms, the scalar potential consists of 39 different terms which we don't write for simplicity. 

Regarding neutrino masses, as we pointed out before there is no tree-level mass term for the neutrinos, since the exotic charges of the new fermions forbid the Standard Model-like coupling with the Higgs. Moreover, note that the Weinberg operator $\bar{L}^c L H H$ is also forbidden by the same $U(1)_{B-L}$ charges. The leading contribution to neutrino masses will arise at the radiative level  coming from the allowed operator $\bar{L}^c LHH \chi$ as shown in Figure~\ref{diagloop}. 

 A rough estimation for neutrino masses coming from the diagram in Figure \ref{diagloop} is given by
\begin{equation} %\label{eq:}
    m_\nu \sim \frac{1}{16\pi^2} Y^2 \lambda \kappa \text{u} \text{v}^2 \frac{1}{\Lambda^3},
\end{equation}
where u is the vev of $\chi$ and $\Lambda$ is the characteristic scale of the loop. The mass of the dark matter candidate will necessarily be lower than this scale. Note that in order to have two massive neutrinos only one generation of $F$ is needed, while two generations of $F$ can generate three non-zero neutrino masses. This is due to, as usual, the sum of two contributions: one coming from the diagram depicted in Figure~\ref{diagloop} and another coming from its transpose.
An estimate of the neutrino mass scale can be obtained if one considers that $\text{u} \sim \mathcal{O}(10)$ GeV. With $\kappa$ order $1$ GeV, $Y \sim \mathcal{O}(0.1)$ and $\lambda \sim \mathcal{O}(1)$, one can fit the atmospheric scale of $0.05$ eV with masses of order $10$ TeV.

Moreover, as can be seen from Figure~\ref{diagloop} all the particles running in the neutrino mass loop are \textit{odd} under the residual $\mathcal{Z}_4$ symmetry. 
Thus, they all belong to the dark sector with the lightest among them, i.e. the lightest out of $\eta_i$ and $F_{L/R}$, being a good candidate for stable dark matter. 
As mentioned before the stability of the dark matter is owing to the fact that all the dark sector particles have charges that are odd under the residual $\mathcal{Z}_4$ symmetry.
However,  all the Standard Model particles are even under $\mathcal{Z}_4$.
Hence, for the lightest dark sector particle there is no possible effective decay operator at any order allowed by the remnant $\mathcal{Z}_4$, see Figure~\ref{U1mod}.

%%%%%%%%%%%%%%%%%%%%%%%%%%%%%%%%%%%%%%%%%%%%%%%%%%%%%%%%%%%%%%%%%%%%%%%%
\section{Summary and conclusion} \label{summary}
%%%%%%%%%%%%%%%%%%%%%%%%%%%%%%%%%%%%%%%%%%%%%%%%%%%%%%%%%%%%%%%%%%%%%%%%

To summarize, neutrino mass and dark matter remain two of the most important shortcomings of the Standard Model. Scotogenic models where the dark sector particles run in the neutrino mass loop provide a particularly attractive scenario to address both these shortcomings in a \sm extension.
In this work, we have shown that the scotogenic symmetry responsible for the dark matter stability can be obtained as a residual $\mathcal{Z}_{2n}$ subgroup of the $U(1)_{B-L}$ symmetry already present in the Standard Model.
We then briefly listed out the general conditions required for our formalism to work for any even residual $\mathcal{Z}_{2n}$ subgroup, previously done for the case of Dirac neutrinos \cite{Bonilla:2018ynb}. We showed that our framework can be applied broadly to many different cases, yet unexplored. Particularizing to a simple case with just one extra scalar Higgs singlet, we discussed all the possible realizations at one-loop level. At the end, one simple realistic example with a remnant $\mathcal{Z}_{4}$ symmetry is explained in more detail to illustrate how the spontaneous symmetry breaking of $U(1)_{B-L}$ to an even $\mathcal{Z}_{2n}$ can be easily accommodated, granting the stability of dark matter.

Before ending we will like to remark that, although in this work we only fleshed out the case for one-loop models but our formalism can be implemented at higher loops and for any even $\mathcal{Z}_{2n}$ symmetry. 
Finally, since in our formalism the global $U(1)_{B-L}$ symmetry is maintained to be anomaly free, therefore all models based on our formalism can be gauged in a straightforward manner. Such gauged models will lead to an even richer phenomenology.

%%%%%%%%%%%%%%%%%%%%%%%%%%%%%%%%%%%%%%%%%%%%%%%%%%%%%%%%%%%%%%%%%%%%%%%%
\section{Ackowledgements}
%%%%%%%%%%%%%%%%%%%%%%%%%%%%%%%%%%%%%%%%%%%%%%%%%%%%%%%%%%%%%%%%%%%%%%%%

EP would like to thank the group AHEP (IFIC)  for the hospitality during his visits. RS  would like to thank IFUNAM for the warm hospitality during his visit.
This work is supported by the Spanish grants SEV-2014-0398, FPA2017-85216-P (AEI/FEDER, UE), Red Consolider MultiDark FPA2017-90566-REDC and PROMETEOII/2014/084 (Generalitat  Valenciana). SCC is supported by the Spanish grant BES-2016-076643. RC is supported by the Spanish grant FPU15/03158. EP is supported in part by DGAPA-PAPIIT IN107118, the German-Mexican research collaboration grant SP 778/4-1 (DFG) and 278017 (CONACyT). 

 \bibliographystyle{utphys}
 \bibliography{bibliography}

\providecommand{\href}[2]{#2}\begingroup\raggedright\begin{thebibliography}{10}

\bibitem{Chatrchyan:2012xdj}
{\bfseries CMS} Collaboration, S.~Chatrchyan {\em et~al.}, ``{Observation of a
  new boson at a mass of 125 GeV with the CMS experiment at the LHC},''
  \href{http://dx.doi.org/10.1016/j.physletb.2012.08.021}{{\em Phys. Lett.}
  {\bfseries B716} (2012) 30--61},
\href{http://arxiv.org/abs/1207.7235}{{\ttfamily arXiv:1207.7235 [hep-ex]}}.
%%CITATION = ARXIV:1207.7235;%%.

\bibitem{Aad:2012tfa}
{\bfseries ATLAS} Collaboration, G.~Aad {\em et~al.}, ``{Observation of a new
  particle in the search for the Standard Model Higgs boson with the ATLAS
  detector at the LHC},''
  \href{http://dx.doi.org/10.1016/j.physletb.2012.08.020}{{\em Phys. Lett.}
  {\bfseries B716} (2012) 1--29},
\href{http://arxiv.org/abs/1207.7214}{{\ttfamily arXiv:1207.7214 [hep-ex]}}.
%%CITATION = ARXIV:1207.7214;%%.

\bibitem{Akerib:2016vxi}
{\bfseries LUX} Collaboration, D.~S. Akerib {\em et~al.}, ``{Results from a
  search for dark matter in the complete LUX exposure},''
  \href{http://dx.doi.org/10.1103/PhysRevLett.118.021303}{{\em Phys. Rev.
  Lett.} {\bfseries 118} no.~2, (2017) 021303},
\href{http://arxiv.org/abs/1608.07648}{{\ttfamily arXiv:1608.07648
  [astro-ph.CO]}}.
%%CITATION = ARXIV:1608.07648;%%.

\bibitem{Aprile:2018dbl}
{\bfseries XENON} Collaboration, E.~Aprile {\em et~al.}, ``{Dark Matter Search
  Results from a One Ton-Year Exposure of XENON1T},''
  \href{http://dx.doi.org/10.1103/PhysRevLett.121.111302}{{\em Phys. Rev.
  Lett.} {\bfseries 121} no.~11, (2018) 111302},
\href{http://arxiv.org/abs/1805.12562}{{\ttfamily arXiv:1805.12562
  [astro-ph.CO]}}.
%%CITATION = ARXIV:1805.12562;%%.

\bibitem{Aghanim:2018eyx}
{\bfseries Planck} Collaboration, N.~Aghanim {\em et~al.}, ``{Planck 2018
  results. VI. Cosmological parameters},''
\href{http://arxiv.org/abs/1807.06209}{{\ttfamily arXiv:1807.06209
  [astro-ph.CO]}}.
%%CITATION = ARXIV:1807.06209;%%.

\bibitem{McDonald:2016ixn}
A.~B. McDonald, ``{Nobel Lecture: The Sudbury Neutrino Observatory: Observation
  of flavor change for solar neutrinos},''
\href{http://dx.doi.org/10.1103/RevModPhys.88.030502}{{\em Rev. Mod. Phys.}
  {\bfseries 88} no.~3, (2016) 030502}.
%%CITATION = RMPHA,88,030502;%%.

\bibitem{Kajita:2016cak}
T.~Kajita, ``{Nobel Lecture: Discovery of atmospheric neutrino oscillations},''
\href{http://dx.doi.org/10.1103/RevModPhys.88.030501}{{\em Rev. Mod. Phys.}
  {\bfseries 88} no.~3, (2016) 030501}.
%%CITATION = RMPHA,88,030501;%%.

\bibitem{An:2012eh}
{\bfseries Daya Bay} Collaboration, F.~P. An {\em et~al.}, ``{Observation of
  electron-antineutrino disappearance at Daya Bay},''
  \href{http://dx.doi.org/10.1103/PhysRevLett.108.171803}{{\em Phys. Rev.
  Lett.} {\bfseries 108} (2012) 171803},
\href{http://arxiv.org/abs/1203.1669}{{\ttfamily arXiv:1203.1669 [hep-ex]}}.
%%CITATION = ARXIV:1203.1669;%%.

\bibitem{Abe:2017uxa}
{\bfseries T2K} Collaboration, K.~Abe {\em et~al.}, ``{Combined Analysis of
  Neutrino and Antineutrino Oscillations at T2K},''
  \href{http://dx.doi.org/10.1103/PhysRevLett.118.151801}{{\em Phys. Rev.
  Lett.} {\bfseries 118} no.~15, (2017) 151801},
\href{http://arxiv.org/abs/1701.00432}{{\ttfamily arXiv:1701.00432 [hep-ex]}}.
%%CITATION = ARXIV:1701.00432;%%.

\bibitem{deSalas:2018bym}
P.~F. De~Salas, S.~Gariazzo, O.~Mena, C.~A. Ternes, and M.~Tórtola,
  ``{Neutrino Mass Ordering from Oscillations and Beyond: 2018 Status and
  Future Prospects},'' \href{http://dx.doi.org/10.3389/fspas.2018.00036}{{\em
  Front. Astron. Space Sci.} {\bfseries 5} (2018) 36},
\href{http://arxiv.org/abs/1806.11051}{{\ttfamily arXiv:1806.11051 [hep-ph]}}.
%%CITATION = ARXIV:1806.11051;%%.

\bibitem{deSalas:2017kay}
P.~F. de~Salas, D.~V. Forero, C.~A. Ternes, M.~Tortola, and J.~W.~F. Valle,
  ``{Status of neutrino oscillations 2018: 3$\sigma$ hint for normal mass
  ordering and improved CP sensitivity},''
  \href{http://dx.doi.org/10.1016/j.physletb.2018.06.019}{{\em Phys. Lett.}
  {\bfseries B782} (2018) 633--640},
\href{http://arxiv.org/abs/1708.01186}{{\ttfamily arXiv:1708.01186 [hep-ph]}}.
%%CITATION = ARXIV:1708.01186;%%.

\bibitem{Ma:1998dn}
E.~Ma, ``{Pathways to naturally small neutrino masses},''
  \href{http://dx.doi.org/10.1103/PhysRevLett.81.1171}{{\em Phys. Rev. Lett.}
  {\bfseries 81} (1998) 1171--1174},
\href{http://arxiv.org/abs/hep-ph/9805219}{{\ttfamily arXiv:hep-ph/9805219
  [hep-ph]}}.
%%CITATION = HEP-PH/9805219;%%.

\bibitem{CentellesChulia:2018gwr}
S.~Centelles~Chuli\'{a}, R.~Srivastava, and J.~W.~F. Valle, ``{Seesaw roadmap
  to neutrino mass and dark matter},''
  \href{http://dx.doi.org/10.1016/j.physletb.2018.03.046}{{\em Phys. Lett.}
  {\bfseries B781} (2018) 122--128},
\href{http://arxiv.org/abs/1802.05722}{{\ttfamily arXiv:1802.05722 [hep-ph]}}.
%%CITATION = ARXIV:1802.05722;%%.

\bibitem{Minkowski:1977sc}
P.~Minkowski, ``{$\mu \to e\gamma$ at a Rate of One Out of $10^{9}$ Muon
  Decays?},''
\href{http://dx.doi.org/10.1016/0370-2693(77)90435-X}{{\em Phys. Lett.}
  {\bfseries 67B} (1977) 421--428}.
%%CITATION = PHLTA,67B,421;%%.

\bibitem{Yanagida:1979as}
T.~Yanagida, ``{Horizontal symmetry and masses of neutrinos},''
{\em Workshop on the baryon number of the Universe and unified theories, O.
  Sawada and A. Sugamoto, eds.} (1979) 95.
%%CITATION = CONFP,C7902131,95;%%.

\bibitem{Mohapatra:1979ia}
R.~N. Mohapatra and G.~Senjanovic, ``{Neutrino mass and spontaneous parity
  violation},''
\href{http://dx.doi.org/10.1103/PhysRevLett.44.912}{{\em Phys. Rev. Lett.}
  {\bfseries 44} (1980) 912}.
%%CITATION = PRLTA,44,912;%%.

\bibitem{Schechter:1981cv}
J.~Schechter and J.~W.~F. Valle, ``{Neutrino Decay and Spontaneous Violation of
  Lepton Number},''
\href{http://dx.doi.org/10.1103/PhysRevD.25.774}{{\em Phys. Rev.} {\bfseries
  D25} (1982) 774}.
%%CITATION = PHRVA,D25,774;%%.

\bibitem{Schechter:1980gr}
J.~Schechter and J.~Valle, ``{Neutrino masses in $SU(2) \times U(1)$
  theories},''
\href{http://dx.doi.org/10.1103/PhysRevD.22.2227}{{\em Phys. Rev.} {\bfseries
  D22} (1980) 2227}.
%%CITATION = PHRVA,D22,2227;%%.

\bibitem{Foot:1988aq}
R.~Foot, H.~Lew, X.~G. He, and G.~C. Joshi, ``{Seesaw Neutrino Masses Induced
  by a Triplet of Leptons},''
\href{http://dx.doi.org/10.1007/BF01415558}{{\em Z. Phys.} {\bfseries C44}
  (1989) 441}.
%%CITATION = ZEPYA,C44,441;%%.

\bibitem{Zee:1980ai}
A.~Zee, ``{A theory of lepton number violation, neutrino Majorana mass, and
  oscillation},''
\href{http://dx.doi.org/10.1016/0370-2693(80)90349-4,
  10.1016/0370-2693(80)90349-4}{{\em Phys.Lett.} {\bfseries B93} (1980) 389}.
%%CITATION = PHLTA,B93,389;%%.

\bibitem{Babu:1988ki}
K.~S. Babu, ``{Model of ``calculable'' Majorana neutrino masses},''
\href{http://dx.doi.org/10.1016/0370-2693(88)91584-5}{{\em Phys. Lett.}
  {\bfseries B203} (1988) 132--136}.
%%CITATION = PHLTA,B203,132;%%.

\bibitem{Ma:2006km}
E.~Ma, ``{Verifiable radiative seesaw mechanism of neutrino mass and dark
  matter},'' \href{http://dx.doi.org/10.1103/PhysRevD.73.077301}{{\em Phys.
  Rev.} {\bfseries D73} (2006) 077301},
\href{http://arxiv.org/abs/hep-ph/0601225}{{\ttfamily arXiv:hep-ph/0601225
  [hep-ph]}}.
%%CITATION = HEP-PH/0601225;%%.

\bibitem{Bonnet:2012kz}
F.~Bonnet, M.~Hirsch, T.~Ota, and W.~Winter, ``{Systematic study of the d=5
  Weinberg operator at one-loop order},''
  \href{http://dx.doi.org/10.1007/JHEP07(2012)153}{{\em JHEP} {\bfseries 07}
  (2012) 153},
\href{http://arxiv.org/abs/1204.5862}{{\ttfamily arXiv:1204.5862 [hep-ph]}}.
%%CITATION = ARXIV:1204.5862;%%.

\bibitem{Ma:2013mga}
E.~Ma, ``{Radiative Origin of All Quark and Lepton Masses through Dark Matter
  with Flavor Symmetry},''
  \href{http://dx.doi.org/10.1103/PhysRevLett.112.091801}{{\em Phys. Rev.
  Lett.} {\bfseries 112} (2014) 091801},
\href{http://arxiv.org/abs/1311.3213}{{\ttfamily arXiv:1311.3213 [hep-ph]}}.
%%CITATION = ARXIV:1311.3213;%%.

\bibitem{Sierra:2014rxa}
D.~Aristizabal~Sierra, A.~Degee, L.~Dorame, and M.~Hirsch, ``{Systematic
  classification of two-loop realizations of the Weinberg operator},''
  \href{http://dx.doi.org/10.1007/JHEP03(2015)040}{{\em JHEP} {\bfseries 03}
  (2015) 040},
\href{http://arxiv.org/abs/1411.7038}{{\ttfamily arXiv:1411.7038 [hep-ph]}}.
%%CITATION = ARXIV:1411.7038;%%.

\bibitem{Ma:2015xla}
E.~Ma, ``{Derivation of Dark Matter Parity from Lepton Parity},''
  \href{http://dx.doi.org/10.1103/PhysRevLett.115.011801}{{\em Phys. Rev.
  Lett.} {\bfseries 115} no.~1, (2015) 011801},
\href{http://arxiv.org/abs/1502.02200}{{\ttfamily arXiv:1502.02200 [hep-ph]}}.
%%CITATION = ARXIV:1502.02200;%%.

\bibitem{CarcamoHernandez:2016pdu}
A.~E. Cárcamo~Hernández, S.~Kovalenko, and I.~Schmidt, ``{Radiatively
  generated hierarchy of lepton and quark masses},''
  \href{http://dx.doi.org/10.1007/JHEP02(2017)125}{{\em JHEP} {\bfseries 02}
  (2017) 125},
\href{http://arxiv.org/abs/1611.09797}{{\ttfamily arXiv:1611.09797 [hep-ph]}}.
%%CITATION = ARXIV:1611.09797;%%.

\bibitem{Cepedello:2017eqf}
R.~Cepedello, M.~Hirsch, and J.~Helo, ``{Loop neutrino masses from $d = 7$
  operator},'' \href{http://dx.doi.org/10.1007/JHEP07(2017)079}{{\em JHEP}
  {\bfseries 1707} (2017) 079},
  \href{http://arxiv.org/abs/1705.01489}{{\ttfamily arXiv:1705.01489
  [hep-ph]}}.

\bibitem{Bernal:2017xat}
N.~Bernal, A.~E. Cárcamo~Hernández, I.~de~Medeiros~Varzielas, and
  S.~Kovalenko, ``{Fermion masses and mixings and dark matter constraints in a
  model with radiative seesaw mechanism},''
  \href{http://dx.doi.org/10.1007/JHEP05(2018)053}{{\em JHEP} {\bfseries 05}
  (2018) 053},
\href{http://arxiv.org/abs/1712.02792}{{\ttfamily arXiv:1712.02792 [hep-ph]}}.
%%CITATION = ARXIV:1712.02792;%%.

\bibitem{Cepedello:2018rfh}
R.~Cepedello, R.~M. Fonseca, and M.~Hirsch, ``{Systematic classification of
  three-loop realizations of the Weinberg operator},''
  \href{http://dx.doi.org/10.1007/JHEP10(2018)197}{{\em JHEP} {\bfseries 1810}
  (2018) 197}, \href{http://arxiv.org/abs/1807.00629}{{\ttfamily
  arXiv:1807.00629 [hep-ph]}}.

\bibitem{Anamiati:2018cuq}
G.~Anamiati, O.~Castillo-Felisola, R.~M. Fonseca, J.~C. Helo, and M.~Hirsch,
  ``{High-dimensional neutrino masses},''
  \href{http://dx.doi.org/10.1007/JHEP12(2018)066}{{\em JHEP} {\bfseries 12}
  (2018) 066},
\href{http://arxiv.org/abs/1806.07264}{{\ttfamily arXiv:1806.07264 [hep-ph]}}.
%%CITATION = ARXIV:1806.07264;%%.

\bibitem{CarcamoHernandez:2019cbd}
A.~E. Cárcamo~Hernández, S.~Kovalenko, R.~Pasechnik, and I.~Schmidt,
  ``{Sequentially loop-generated quark and lepton mass hierarchies in an
  extended Inert Higgs Doublet model},''
\href{http://arxiv.org/abs/1901.02764}{{\ttfamily arXiv:1901.02764 [hep-ph]}}.
%%CITATION = ARXIV:1901.02764;%%.

\bibitem{Hirsch:2017col}
M.~Hirsch, R.~Srivastava, and J.~W.~F. Valle, ``{Can one ever prove that
  neutrinos are Dirac particles?},''
  \href{http://dx.doi.org/10.1016/j.physletb.2018.03.073}{{\em Phys. Lett.}
  {\bfseries B781} (2018) 302--305},
\href{http://arxiv.org/abs/1711.06181}{{\ttfamily arXiv:1711.06181 [hep-ph]}}.
%%CITATION = ARXIV:1711.06181;%%.

\bibitem{Bonilla:2018ynb}
C.~Bonilla, S.~Centelles~Chuli\'{a}, R.~Cepedello, E.~Peinado, and
  R.~Srivastava, ``{Dark matter stability and Dirac neutrinos using only
  Standard Model symmetries},''
\href{http://arxiv.org/abs/1812.01599}{{\ttfamily arXiv:1812.01599 [hep-ph]}}.
%%CITATION = ARXIV:1812.01599;%%.

\bibitem{Chulia:2016ngi}
S.~Centelles~Chuliá, E.~Ma, R.~Srivastava, and J.~W.~F. Valle, ``{Dirac
  Neutrinos and Dark Matter Stability from Lepton Quarticity},''
  \href{http://dx.doi.org/10.1016/j.physletb.2017.01.070}{{\em Phys. Lett.}
  {\bfseries B767} (2017) 209--213},
\href{http://arxiv.org/abs/1606.04543}{{\ttfamily arXiv:1606.04543 [hep-ph]}}.
%%CITATION = ARXIV:1606.04543;%%.

\bibitem{Montero:2007cd}
J.~C. Montero and V.~Pleitez, ``{Gauging U(1) symmetries and the number of
  right-handed neutrinos},''
  \href{http://dx.doi.org/10.1016/j.physletb.2009.03.065}{{\em Phys. Lett.}
  {\bfseries B675} (2009) 64--68},
\href{http://arxiv.org/abs/0706.0473}{{\ttfamily arXiv:0706.0473 [hep-ph]}}.
%%CITATION = ARXIV:0706.0473;%%.

\bibitem{Ma:2014qra}
E.~Ma and R.~Srivastava, ``{Dirac or inverse seesaw neutrino masses with $B-L$
  gauge symmetry and $S_3$ flavor symmetry},''
  \href{http://dx.doi.org/10.1016/j.physletb.2014.12.049}{{\em Phys. Lett.}
  {\bfseries B741} (2015) 217--222},
\href{http://arxiv.org/abs/1411.5042}{{\ttfamily arXiv:1411.5042 [hep-ph]}}.
%%CITATION = ARXIV:1411.5042;%%.

\bibitem{Ma:2015mjd}
E.~Ma, N.~Pollard, R.~Srivastava, and M.~Zakeri, ``{Gauge $B-L$ Model with
  Residual $Z_3$ Symmetry},''
  \href{http://dx.doi.org/10.1016/j.physletb.2015.09.010}{{\em Phys. Lett.}
  {\bfseries B750} (2015) 135--138},
\href{http://arxiv.org/abs/1507.03943}{{\ttfamily arXiv:1507.03943 [hep-ph]}}.
%%CITATION = ARXIV:1507.03943;%%.

\bibitem{Ma:2015raa}
E.~Ma and R.~Srivastava, ``{Dirac or inverse seesaw neutrino masses from gauged
  $B–L$ symmetry},'' \href{http://dx.doi.org/10.1142/S0217732315300207}{{\em
  Mod. Phys. Lett.} {\bfseries A30} no.~26, (2015) 1530020},
\href{http://arxiv.org/abs/1504.00111}{{\ttfamily arXiv:1504.00111 [hep-ph]}}.
%%CITATION = ARXIV:1504.00111;%%.

\bibitem{Ho:2016aye}
S.-Y. Ho, T.~Toma, and K.~Tsumura, ``{Systematic $U(1)_{B–L}$ extensions of
  loop-induced neutrino mass models with dark matter},''
  \href{http://dx.doi.org/10.1103/PhysRevD.94.033007}{{\em Phys. Rev.}
  {\bfseries D94} no.~3, (2016) 033007},
\href{http://arxiv.org/abs/1604.07894}{{\ttfamily arXiv:1604.07894 [hep-ph]}}.
%%CITATION = ARXIV:1604.07894;%%.

\bibitem{Ma:2016nnn}
E.~Ma, N.~Pollard, O.~Popov, and M.~Zakeri, ``{Gauge $B–L$ model of radiative
  neutrino mass with multipartite dark matter},''
  \href{http://dx.doi.org/10.1142/S0217732316501637}{{\em Mod. Phys. Lett.}
  {\bfseries A31} no.~27, (2016) 1650163},
\href{http://arxiv.org/abs/1605.00991}{{\ttfamily arXiv:1605.00991 [hep-ph]}}.
%%CITATION = ARXIV:1605.00991;%%.

\bibitem{Patra:2016ofq}
S.~Patra, W.~Rodejohann, and C.~E. Yaguna, ``{A new B - L model without
  right-handed neutrinos},''
  \href{http://dx.doi.org/10.1007/JHEP09(2016)076}{{\em JHEP} {\bfseries 09}
  (2016) 076},
\href{http://arxiv.org/abs/1607.04029}{{\ttfamily arXiv:1607.04029 [hep-ph]}}.
%%CITATION = ARXIV:1607.04029;%%.

\bibitem{Wang:2016lve}
W.~Wang and Z.-L. Han, ``{Naturally Small Dirac Neutrino Mass with Intermediate
  $SU(2)_{L}$ Multiplet Fields},''
  \href{http://arxiv.org/abs/1611.03240}{{\ttfamily arXiv:1611.03240
  [hep-ph]}}.
[JHEP04,166(2017)].
%%CITATION = ARXIV:1611.03240;%%.

\bibitem{Wang:2017mcy}
W.~Wang, R.~Wang, Z.-L. Han, and J.-Z. Han, ``{The $B-L$ Scotogenic Models for
  Dirac Neutrino Masses},''
  \href{http://dx.doi.org/10.1140/epjc/s10052-017-5446-9}{{\em Eur. Phys. J.}
  {\bfseries C77} no.~12, (2017) 889},
\href{http://arxiv.org/abs/1705.00414}{{\ttfamily arXiv:1705.00414 [hep-ph]}}.
%%CITATION = ARXIV:1705.00414;%%.

\bibitem{Nanda:2017bmi}
D.~Nanda and D.~Borah, ``{Common origin of neutrino mass and dark matter from
  anomaly cancellation requirements of a $U(1)_{B-L}$ model},''
  \href{http://dx.doi.org/10.1103/PhysRevD.96.115014}{{\em Phys. Rev.}
  {\bfseries D96} no.~11, (2017) 115014},
\href{http://arxiv.org/abs/1709.08417}{{\ttfamily arXiv:1709.08417 [hep-ph]}}.
%%CITATION = ARXIV:1709.08417;%%.

\bibitem{Han:2018zcn}
Z.-L. Han and W.~Wang, ``{$Z'$ Portal Dark Matter in $B-L$ Scotogenic Dirac
  Model},'' \href{http://dx.doi.org/10.1140/epjc/s10052-018-6308-9}{{\em Eur.
  Phys. J.} {\bfseries C78} no.~10, (2018) 839},
\href{http://arxiv.org/abs/1805.02025}{{\ttfamily arXiv:1805.02025 [hep-ph]}}.
%%CITATION = ARXIV:1805.02025;%%.

\bibitem{Kang:2018lyy}
S.~K. Kang and O.~Popov, ``{Radiative neutrino mass via fermion kinetic
  mixing},'' \href{http://dx.doi.org/10.1103/PhysRevD.98.115025}{{\em Phys.
  Rev.} {\bfseries D98} no.~11, (2018) 115025},
\href{http://arxiv.org/abs/1807.07988}{{\ttfamily arXiv:1807.07988 [hep-ph]}}.
%%CITATION = ARXIV:1807.07988;%%.

\bibitem{Modak:2016ung}
T.~Modak, S.~Sadhukhan, and R.~Srivastava, ``{750 GeV diphoton excess from
  gauged $B - L$ symmetry},''
  \href{http://dx.doi.org/10.1016/j.physletb.2016.03.021}{{\em Phys. Lett.}
  {\bfseries B756} (2016) 405--412},
\href{http://arxiv.org/abs/1601.00836}{{\ttfamily arXiv:1601.00836 [hep-ph]}}.
%%CITATION = ARXIV:1601.00836;%%.

\bibitem{Singirala:2017see}
S.~Singirala, R.~Mohanta, and S.~Patra, ``{Singlet scalar Dark matter in
  $U(1)_{B-L}$ models without right-handed neutrinos},''
  \href{http://dx.doi.org/10.1140/epjp/i2018-12270-0}{{\em Eur. Phys. J. Plus}
  {\bfseries 133} no.~11, (2018) 477},
\href{http://arxiv.org/abs/1704.01107}{{\ttfamily arXiv:1704.01107 [hep-ph]}}.
%%CITATION = ARXIV:1704.01107;%%.

\bibitem{DeRomeri:2017oxa}
V.~De~Romeri, E.~Fernandez-Martinez, J.~Gehrlein, P.~A.~N. Machado, and
  V.~Niro, ``{Dark Matter and the elusive $Z'$ in a dynamical Inverse Seesaw
  scenario},'' \href{http://dx.doi.org/10.1007/JHEP10(2017)169}{{\em JHEP}
  {\bfseries 10} (2017) 169},
\href{http://arxiv.org/abs/1707.08606}{{\ttfamily arXiv:1707.08606 [hep-ph]}}.
%%CITATION = ARXIV:1707.08606;%%.

\bibitem{Nomura:2017jxb}
T.~Nomura and H.~Okada, ``{Neutrinophilic two Higgs doublet model with dark
  matter under an alternative $U(1)_{B-L}$ gauge symmetry},''
  \href{http://dx.doi.org/10.1140/epjc/s10052-018-5667-6}{{\em Eur. Phys. J.}
  {\bfseries C78} no.~3, (2018) 189},
\href{http://arxiv.org/abs/1708.08737}{{\ttfamily arXiv:1708.08737 [hep-ph]}}.
%%CITATION = ARXIV:1708.08737;%%.

\bibitem{Das:2017deo}
A.~Das, N.~Okada, and D.~Raut, ``{Heavy Majorana neutrino pair productions at
  the LHC in minimal U(1) extended Standard Model},''
  \href{http://dx.doi.org/10.1140/epjc/s10052-018-6171-8}{{\em Eur. Phys. J.}
  {\bfseries C78} no.~9, (2018) 696},
\href{http://arxiv.org/abs/1711.09896}{{\ttfamily arXiv:1711.09896 [hep-ph]}}.
%%CITATION = ARXIV:1711.09896;%%.

\bibitem{Okada:2018tgy}
N.~Okada, S.~Okada, and D.~Raut, ``{A natural $Z^\prime$-portal Majorana dark
  matter in alternative U(1) extended Standard Model},''
\href{http://arxiv.org/abs/1811.11927}{{\ttfamily arXiv:1811.11927 [hep-ph]}}.
%%CITATION = ARXIV:1811.11927;%%.

\bibitem{Calle:2018ovc}
J.~Calle, D.~Restrepo, C.~E. Yaguna, and O.~Zapata, ``{Minimal radiative Dirac
  neutrino mass models},''
\href{http://arxiv.org/abs/1812.05523}{{\ttfamily arXiv:1812.05523 [hep-ph]}}.
%%CITATION = ARXIV:1812.05523;%%.

\end{thebibliography}\endgroup

\end{document}